\documentclass[%
prl,
reprint,
groupedaddress,
 amsmath,amssymb,
 aps,
]{revtex4-2}

\usepackage{physics}
\usepackage{graphicx}
\usepackage{dcolumn}
\usepackage{bm}
\usepackage{hyperref}
\usepackage{xcolor}

\renewcommand\vec{\bm}
\newcommand{\Z}{\ensuremath{\mathbb{Z}_2}}
\newcommand{\tmix}{\ensuremath{t_{\rm{mix}}}}
\newcommand{\tres}{\ensuremath{t_{\rm{res}}}}

\def\equationautorefname~#1\null{Eq. (#1)\null}

\newcommand{\bs}{\boldsymbol{s}}

\begin{document}

\title{Gibbs resampling transitions and structured fast-measurement protocols}

\author{Daan Timmers}
\email{daan.timmers@physics.ox.ac.uk}

\author{Benedikt Placke}
\author{S. A. Parameswaran}

\affiliation{
 Rudolf Peierls Centre for Theoretical Physics, Clarendon Laboratory, Parks Road, Oxford, OX1 3PU, UK
}

\begin{abstract}
Quantum Gibbs sampling algorithms generalize classical Markov chain Monte Carlo (cMCMC) methods, and prepare thermal states on quantum computers by implementing local dissipative evolution for a mixing time $t_{\rm mix}$. Since quantum measurements disrupt the prepared state, the rate of extracting unbiased information from the system depends on the recovery of the underlying thermal correlations. The corresponding ``Gibbs resampling'' time $t_{\rm res}$ is distinct both from $t_{\rm mix}$ and from the autocorrelation time relevant to cMCMC sampling. We study this problem in effective classical models that emulate disruptive quantum measurements of local observables on extensive subsystems. When such subsystems are chosen randomly, in models with non-local order parameters we uncover a ``resampling transition'' as a function of the fraction of sites $p$ that are measured, from $t_{\rm res}\propto \log L$ for small $p$ to $t_{\rm res}\propto {\rm poly} L$ for large $p$, in systems of linear size $L$.  We argue that this transition is generically absent in systems with local order parameters, and relate this dichotomy to one between the coarsening of initial states that break 1- and 0-form symmetries under Glauber dynamics. Finally, we devise a structured fast-measurement protocol that evades slow resampling even with non-local order parameters, while leaving only a vanishing fraction of all sites unmeasured.
\end{abstract}

\maketitle

\noindent\textit{Introduction.---} Quantum simulation of many-body phenomena is a significant near-term application of quantum computing platforms that holds the promise of practical quantum advantage~\cite{daley2022PQA,hoefler2023disentanglinghypepracticalityrealistically,babbush2025grandchallengequantumapplications}. Often, an important ingredient of such simulations is the task of preparing a quantum thermal state~\cite{Terhal2000Equilibration,Temme2011QuantumMetropolis,Poulin2009GibbsSampling, Kastoryano2016CommutingGibbs}, for which several recent works have developed  provably efficient approaches~\cite{chen2025_quantum_thermal_sim, ding2025efficient} often discussed under the umbrella term of ``Gibbs sampling''. These methods implement a local dissipative Lindblad dynamics whose stationary state is  the quantum density matrix $\rho_\beta \propto e^{-\beta H}$ of a local many-body Hamiltonian $H$, thereby  mirroring the preparation of a classical Gibbs state as the fixed point of a Markov chain in Monte Carlo (MC) sampling. The timescale for preparing such states is controlled by the mixing time of the Lindbladian~\cite{LevinPeresWilmer2006}. This is governed by physical aspects of the system: if the low-temperature state breaks a symmetry of $H$, the relevant time scale is that of ``phase ordering'' \cite{bray_theory_1994}, or ``mixing within a symmetry sector'' \cite{gheissari_sinclair2022mixing_within_a_phase, bergamaschi2025rapid_within_logical}, and can diverge polynomially in system size. In other cases, e.g. spin glasses, mixing times may even diverge exponentially.
\begin{figure}[t!]
    \centering
    \includegraphics[width=\linewidth]{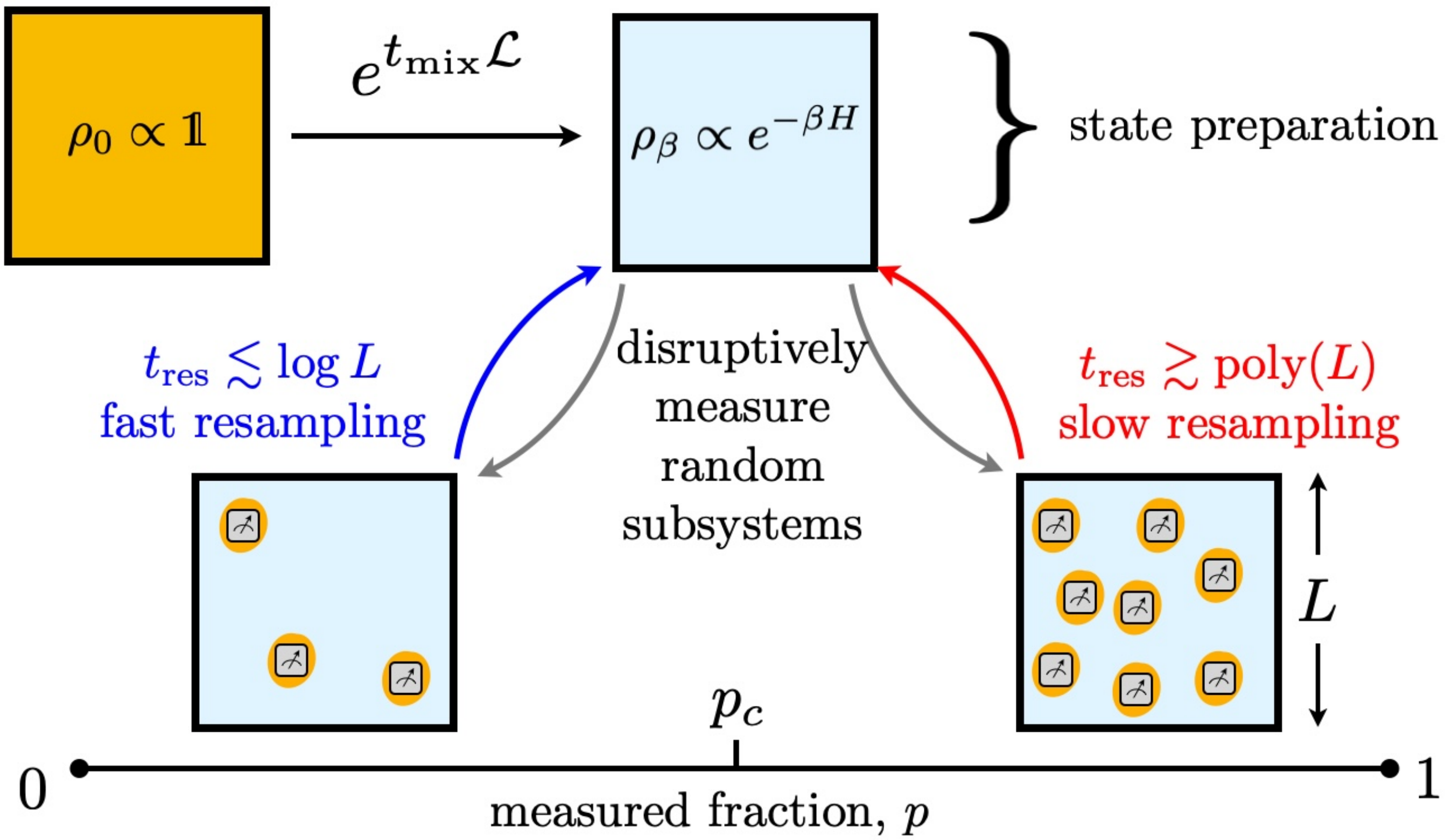}
    \caption{\textbf{``Gibbs Resampling''.} A specified quantum state (e.g. a thermal state of a local Hamiltonian) can be prepared from an arbitrary input using a local Lindbladian $\mathcal{L}$ in the mixing time $\tmix$. A  disruptive measurement is classically modeled by resetting a fraction $p$ of the system to a random state.  The local dynamics then  recovers the initial ordered state in the ``resampling time'' $\tres(p)$. The scaling of $\tres(p)$ with system size $L$ may undergo a transition as $p$ is varied.
    }
    \label{fig:setup}
\end{figure}

We consider related but distinct questions: given such a protocol, how efficiently can the target state be {\it interrogated} by measurements, e.g. to extract local observables, once prepared? How does this depend on the nature of the state under study? The answers directly impact the quality of the corresponding quantum simulation, since it defines the resources necessary to achieve results of a specified precision. In classical MC sampling, the answer is given by the equilibrium autocorrelation time. In the quantum setting however, measurements generically disrupt the state (although  recent work suggests this can be partially mitigated \cite{jiang_predicting_2026,chen2026thermal_expectation,chen_catalytic_2025} at the cost of complicating the measurement process). The open question is then to determine the cost of ``resampling'' the equilibrium state, i.e., of recovering it from the post-measurement state autonomously via the local dissipative dynamics.

Although this ``resampling time'' is bounded above by the mixing time (which governs the time to prepare the target state from {\it any} input) in many cases this is an overestimate: a weak local perturbation of a phase with some bulk order likely heals much more quickly. This intuition is borne out by rigorous results that show that Gibbs samplers rapidly restore fixed-point states that are perturbed only inside a region whose size remains fixed in the thermodynamic limit \cite{chen_quantum_2025}. For states with exponentially decaying correlations, one might expect to do even better by performing simultaneous measurements {\it extensively} across the system, yet doing so sufficiently sparsely that the system retains sufficient memory of the original state to heal quickly. Understanding Gibbs resampling under such extensive protocols and devising adaptive speedups by exploiting the physical properties of the target state can significantly enhance the utility of quantum platforms to probe correlated systems.

Addressing these questions in full generality is a formidable challenge given the complexity of quantum many-body dynamics. Accordingly, in the present work we introduce and explore a classical Markov-chain counterpart of the ``Gibbs resampling'' problem (\autoref{fig:setup}). We focus  on two prototypical types of order, corresponding to low-temperature Gibbs states that spontaneously break symmetries and hence possess a local order parameter, and those that are instead characterized by topological order encoded in non-local correlations. (In modern parlance, these  respectively  break 0-form and 1-form symmetries.) We  model the disruption introduced by quantum measurement in this classical setting by resetting measured degrees of freedom (spins)  to an arbitrary state drawn from the infinite-temperature Gibbs measure, and use classical detailed-balanced Glauber dynamics as a proxy for quantum Gibbs sampling. 
First, we consider spatially uncorrelated measurement protocols that interrogate each spin independently with probability $p$. Remarkably, we find that broken-symmetry Gibbs states characterized by a nonzero value of a local order parameter can be resampled ``rapidly'' (i.e., the Gibbs state is recovered in $t_{\rm res} \sim O(\log L)$ where $L$ is the linear system size), for {\it any} $p<1$ --- even though such states take $t_{\rm mix} \sim {\rm poly}(L)$ to prepare. Specifically, we present  numerical evidence of this for the studied models and give an argument that the result generalizes to any setting with broken 0-form (global) symmetries. In contrast, we find that discrete gauge theories exhibit a ``resampling transition'' at a finite $p_c<1$ for any local dynamics. We explain these observations by combining ideas from classical coarsening and  error correction. Equipped with this understanding, we construct structured measurement protocols for which rapid resampling is possible for both types of order. Finally, we comment on how these classical results might lift to the quantum setting.

\noindent\textit{Models.---} We study models of continuous or discrete spins $\bs_i$ (with $\bs_i^2=1$) on the sites or links of a hypercubic lattice of linear dimension $L$ with a local classical energy functional $\mathcal{H}[\{\bs_i\}]$. We are interested in two  classes of model. The first describes a classical ferromagnet, $\mathcal{H_{\rm FM}}= - \sum_{\langle i, j\rangle} \bs_i \cdot\bs_j$  where the sum is over nearest neighbors. 
Concretely, we will consider Ising spins ($\bs_i=\pm 1$) in dimension $d=2$, and  XY ($\bs_i \in S^1$) spins in $d=3$ (we also discuss $d=3$ Heisenberg spins ($\bs_i \in S^2$) in the Supplementary Material \cite{supmat}). These models have a robust low temperature phase exhibiting long-range spontaneous symmetry breaking order, captured by a local order parameter $m = \frac{1}{N} \sum_i \langle\bs_i\rangle$  where $N=L^d$ is the total number of spins. We will be concerned with the disruption of the ordered state at $T=0$ (which has maximal magnetization $m=1$) by the introduction of topological defects of the order through our measurements. In the Ising case, these are domain walls, while in the 3D XY setting, they are closed vortex lines in the vector order parameter.

We will contrast the behavior of ordered ferromagnets  against that of the classical 3D $\mathbb{Z}_2$ lattice gauge theory (or Wegner's $M_{3,2}$ Ising model \cite{wegner_duality_1971}), $\mathcal{H}_{\rm IGT} = - \sum_{\square} \prod_{i\in \square} \bs_i$, where the sum is over all plaquettes of a cubic lattice with  $N=3L^d$ spins on the links; details on its properties are supplied in the End Matter. The disruption of the ground state by our measurements introduces energetic defects. These are collections of violated plaquette terms, which form closed $\mathbb{Z}_2$ flux loops on the dual lattice. Although the geometry of these flux loops is reminiscent of 3D XY vortex lines, there are intrinsic differences between the two. Most importantly, the circulation of the XY order parameter around a closed loop exactly measures the total vortex charge piercing the enclosed surface, whereas  $\mathbb{Z}_2$ fluxes in the gauge  theory  are detected by  Wilson loops that are intrinsically nonlocal and cannot be related to integrals of a local order parameter.
\begin{figure*}[t!]
    \centering
    \includegraphics[width=\linewidth]{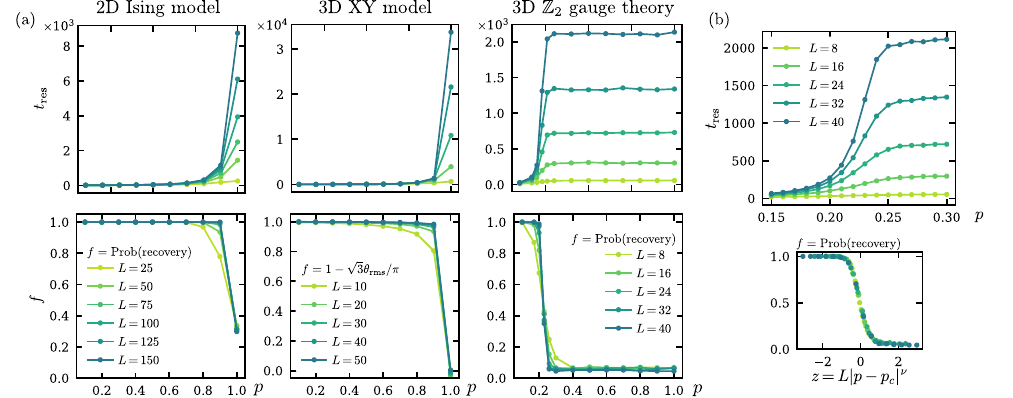}
    \caption{(a) A comparison of resampling time and logical fidelity in the 2D Ising model, the 3D XY model, and 3D classical $\Z$ gauge theory, as a function of measurement probability $p$. In the two models with a local order parameter, the finite size scaling indicates that resampling is rapid, with fidelity close to unity, for any $p<1$, while for the gauge theory there is a resampling transition. (b) A closer look at the resampling transition. Above: resampling time as a function of $p$ near the transition. Below: scaling collapse of logical fidelity. We assume $\nu = 1$, and find $p_c \approx 0.226$.}
    \label{fig:overview}
\end{figure*}

\noindent\textit{Dynamics and measurement protocols.---} We initialize the system in the target ``all up'' configuration which is a fixed point of the $T=0$ Glauber dynamics, and is characterized by maximal magnetization $m=1$ in the ferromagnet, and by the absence of flux loops in IGT. We model a disruptive measurement by attempting to measure each spin with probability $p$; if measured, the spin is ``reset'' to a random configuration. We then evolve the system under Glauber dynamics \cite{glauber_time-dependent_1963} (``model A''~\cite{hohenberg_theory_1977}).

We  define the ``resampling time'' to be the time to return to the ground state in the two discrete models, and to reach $m > 0.98$ in the XY model (reaching $m=1$ exactly takes infinitely long), and compute its average over a large number of independent runs. We also track the ``logical fidelity'' to the initial state. In the discrete models, this is captured by the fraction $f$ of runs that return to the initial state (up to gauge transformations in the case of the IGT), and for continuous spins, by $f = 1-  \theta_{\rm rms} /\theta^0_{\rm rms}$, where $\theta_{\rm rms}$ is the root-mean-squared angular deviation of the magnetization vector at the resampling time from its initial value, normalized to the uniform distribution to give $f=0$ at $p=1$. 

\noindent\textit{Numerical results.---} For the described protocol, in which the measurement-resets are {\it uncorrelated} in space, we  find that the behavior of the resampling time as a function of $p$ is strikingly different depending on whether or not the low-$T$ target state is characterized by a local order parameter (\autoref{fig:overview}). In the Ising and XY cases, where this is the case, the resampling time remains small with system-size dependence absent or very weak, {\it except} at $p=1$, where it shows strong system-size-dependence~\cite{supmat}. Coincident with this, the logical fidelity remains nearly $1$ except near $p=1$. In both cases the finite-size-scaling indicates that $p=1$ is a singular point, with any $p<1$ characterized by rapid resampling to the original state with perfect fidelity. In contrast, the 3D $\mathbb{Z}_2$ gauge theory shows a clear transition in both resampling time and fidelity at a finite $p_c\approx 0.226$. We now rationalize these results using ideas from coarsening and error correction.

\noindent\textit{Rapid resampling with a local order parameter.---} 
To understand the resampling dynamics we use ideas from the theory of phase ordering kinetics, the description of the dynamical emergence of long-range order following a quench from a paramagnet into an ordered phase~\cite{bray_theory_1994}. This ``coarsening dynamics'' is characterized by a single length scale $L(t)$, setting the scale on which the system is ordered at time $t$. For the ``model A'' dynamics studied here, $L(t) \sim t^{1/2}$ \cite{bray_growth_1994}. A finite system of size $L$ becomes fully ordered when $L(t) \sim L$, so that $\tmix \sim L^2$ (within a symmetry sector). The $p=1$ limit of our disruptive measurement protocol is equivalent to a quench from infinite temperature, giving $\tres (p = 1) = \tmix \sim L^2$. The coarsening problem is best understood by describing it as a process of motion and annihilation of topological defects~\cite{bray_theory_1994}. For concreteness, we first consider the Ising model in arbitrary $d \geq 2$, where the defects are domain walls. In coarsening, domain walls drift with a velocity proportional to the local curvature, so that a contractible domain wall of size $R$ shrinks and vanishes in a time $t \sim R^2$ \cite{allen_microscopic_1979}. The description in terms of defects is only appropriate once the domain walls are smooth, i.e., after evolving for some finite time $t_\ell$ so that $L(t_\ell) = \ell \gg \xi$, the equilibrium correlation length. To understand the resampling problem for $p<1$, we need to determine how the domain wall distribution at $t_\ell$ depends on $p$.

First, we consider coarse-graining the initial state (after the disruptive measurement) by averaging over blocks of linear size $\ell$. When $p<1$, the magnetization of each block has a deterministic contribution from the approximately $(1-p) \ell^d$ unmeasured spins, and a stochastic distribution from the approximately $p \ell^d$ spins which were measured. This results in a magnetization density 
\begin{equation}
    \label{eq:ising average magnetization}
    m(p, \ell) = 1 - p + \delta m (p, \ell),
\end{equation}
where, for sufficiently large $\ell$, $\delta m$ can be approximated by a Gaussian random variable with zero mean and standard deviation $\sqrt{p / \ell^d}$. For $\ell \gg \left( \frac{p}{(1-p)^2} \right) ^{1/d}$ the sign of the magnetization of each block is with high probability aligned with that of the all-up initial state. The point $p=1$ is special: the deterministic contribution vanishes and $m(p=1, \ell)$ has a random sign on any length scale.

We now make the reasonable assumption that the microscopic bias of the initial state translates into an equivalent bias of the order parameter field at late times $t > t_\ell$, i.e., that a block of spins which is (weakly) aligned with the all-up state at $t=0$ with high probability evolves to have a large positive magnetization at $t = t_\ell$. We can then use the coarse-grained structure of the initial state to constrain the size of minority domains at late times. The size distribution of minority domains is estimated by considering the site percolation problem on the effective lattice formed by the blocks of size $\ell$, with the relevant occupation probability being the probability that a given block is aligned \emph{opposite} to the all-up state at $t_\ell$. For $\ell$ sufficiently large this probability is small, so that the site percolation problem is well below its threshold. Typical minority domains therefore contain an $O(1)$ number of spins, while the largest minority domain contains $O(\log L)$ spins \cite{grimmett_percolation_1999}. Note that, through the required $\ell$, these domain sizes will also depend on $p$, and will grow large (though not in the scaling sense) for $p$ close to unity.

The resampling time is obtained from the size distribution of the minority domains. As the largest domain contains $O(\log L)$ spins, it has linear size $R \sim O((\log L)^{1/d})$ and hence decays in a time $t \sim O((\log L)^{2/d})$. System size dependence of the resampling time also stems from the fact that there are extensively many minority domains: while on average they decay in a finite time, the time required for \emph{all} domains to have decayed is logarithmic in their number. This contribution scales at least as strongly as the contribution from the largest domain. Therefore, we expect that the Ising model in $d \geq 2$ spatial dimensions has $\tres \sim O(\log L)$, for \emph{any} $p<1$. (This $\log L$ dependence of $t_{\rm res}$ corresponds to \emph{exactly} recovering the target state; the magnetization density will be arbitrarily close to the equilibrium value already in finite time. This is consistent with Ref.~\onlinecite{filipe_phase-ordering_1995}, which approximately solves the time-dependent Ginzburg-Landau dynamics for scalar and vector order parameters, and concludes that these saturate in finite time in the thermodynamic limit when initially non-zero on average.)

A similar argument holds in the 3D XY case, although here minority spin domains are no longer well-defined and the description must be entirely in terms of topological defects. In the End Matter, we use a coarse-graining argument similar to the above to bound the size of defects in $O(n)$ models, and hence argue that resampling is rapid for any $p<1$. The local order parameter $m$ is central to this argument: since the topological charge can be written as an integral over  $m$, which is in turn biased by the initial condition, defects must be confined to small regions. While we only discuss ferromagnetic spin models, we conjecture that rapid resampling is a generic property of any model with a local order parameter.

\noindent\textit{Rapid-slow resampling transition in Ising gauge theory with uncorrelated measurements.---} 
We now explore the converse question --- i.e., whether slow resampling is allowed absent a local order parameter --- by considering 3D classical $\Z$ gauge theory. We can see this must have a resampling transition by viewing  Glauber dynamics as classical error correction protocol. Since measured spins are reset to either up or down with equal probability, a measurement with probability $p$ is equivalent to a spin-flip error with probability $p/2$. The logical fidelity measures the ability of the Glauber dynamics to correct these errors and recover the information stored in the initial state. As an error correction problem, the decoding of 3D classical $\Z$ gauge theory maps exactly to the decoding of the 3D toric code with $X$ noise, which is known to have an optimal threshold of $p_{\rm X} \approx 0.233$ \cite{xu_phenomenological_2025}. In the classical gauge theory, this means that no protocol can recover the initial state when $p>2p_{\rm X}$. We must therefore have a resampling transition, with a threshold $p_c < 0.47$. Numerically  we find $p_c \approx 0.226$, substantially lower than the bound [\autoref{fig:overview}(b)]. For $p<p_c$, resampling is rapid and $f\to 1$ in the thermodynamic limit. Above $p_c$, $t_{\rm res}\propto L^2$, and the fidelity is small. 

From the perspective of coarsening, there is no essential difference between the flux loop defects in the gauge theory, and the loop-like defects in the 2D Ising and 3D XY models. Their late-time dynamics is curvature-driven, and flux loops decay in a time quadratic in their diameter. The difference in behavior between the gauge theory and the local order parameter models is thus not due to distinct late-time scaling, but rather stems from the different distribution of defects that emerges from the initial condition after early-time coarse-graining. The local bias of the initial condition in models with a local order parameter forces the defects to ``cluster'' in confined regions: a restriction absent in the gauge theory.

With hindsight, rapid resampling in the Ising model may also be linked to error correction. At $T=0$ the Ising state space realizes a  repetition code, which has an optimal threshold under bitflip noise of $p_{\rm flip} = \frac{1}{2}$ \cite{nielsen_quantum_2010}, corresponding  to $p=1$ in our protocol;  somewhat surprisingly, local Glauber dynamics  achieves optimal decoding.

\begin{figure}
    \centering
    \includegraphics[width=\linewidth]{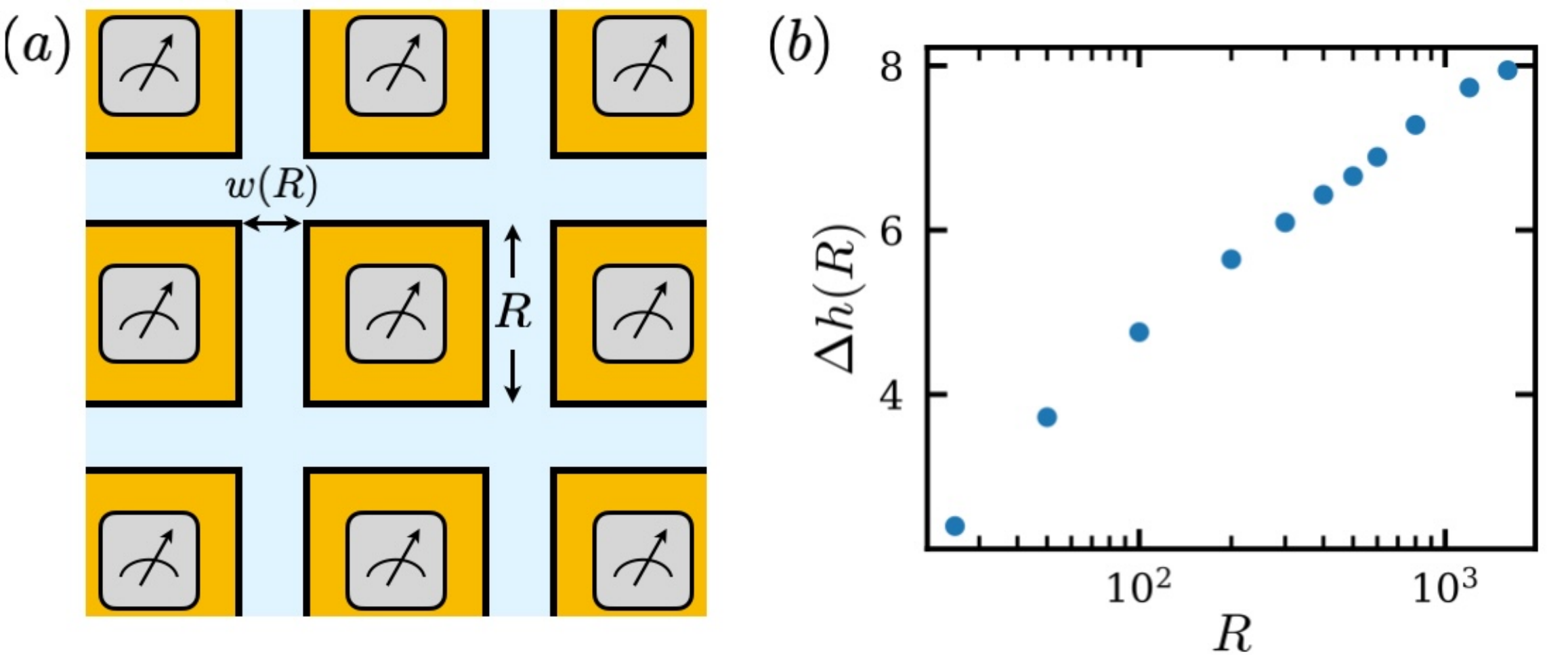}
    \caption{(a) A structured fast-measurement protocol: all spins in boxes of size $R$ are measured, while keeping spins in a ``collar'' of width $w(R)$ around every box unmeasured. (b) Maximum excursion as a function of box size in the 2D Ising model evolving under Glauber dynamics at $T \approx 0.74 \, T_c$.}
    \label{fig:collar construction}
\end{figure}

\noindent\textit{Structured fast-measurement protocols.---} 
While the example of classical $\Z$ gauge theory shows that it is not in general possible to make uncorrelated measurements of an arbitrarily large fraction of spins without paying the cost of slow resampling, this may be circumvented by instead using a deterministic, spatially structured measurement protocol [\autoref{fig:collar construction}(a)]. Here, the system is subdivided into boxes of finite linear size $R$, separated by a ``collar'' of width $w(R)$. Measuring all spins inside each box while leaving those in the collar region unmeasured results in measuring a fraction $1-\epsilon$ of spins, with $\epsilon \sim w(R) / R$.

The construction ensures that measurement-induced topological defects are confined to the boxes, provided the $w(R)$ is chosen large enough to suppress the probability of defects from adjacent boxes fusing after ``excursions'' into the collar. The argument relies only on defect geometry and does not require a local order parameter, and so we can study the Ising case and transfer our results to the IGT. Extended defects, when confined to a box, are convex near its boundary, so that the required $w(R)$ is set by the net effect of fluctuations \emph{against} the inward force due to curvature. For $T=0$ Glauber dynamics there are no fluctuations, so a minimal collar width of $w(R) = 2$ suffices. At finite $T < T_c$, for models with extended topological defects, the required $w(R)$ can bounded above using the \emph{wandering exponent} $\zeta$ of the system of interest, which governs fluctuations \emph{without} curvature. It is defined by the dependence of the equilibrium fluctuations  of a flat defect on its length $R$, as $\Delta h \sim R^\zeta$ \cite{chaikin_principles_1995}. We expect this bound to be very loose \cite{supmat}: e.g., for the 2D Ising model, we find numerically that $\Delta h(R) \sim O(\log R)$ [\autoref{fig:collar construction}(b)], even though $\zeta = \frac{1}{2}$. 

Provided the collar width is sufficient, the spins in each box re-order  in a finite time $t \sim R^2$ on average. Requiring \emph{all} extensively many boxes to re-order yields a resampling time $\tres \sim O(\log L)$. The collar ensures that the logical fidelity is 1. As long as the required collar width $w(R)$ scales sub-linearly with $R$, the fraction of measured spins $1- \epsilon$ can be made arbitrarily large at finite $R$. For the models with a local order parameter, this construction does not offer any advantages either in resampling time or in logical fidelity, when compared to the uncorrelated protocol. In fact, the argument for rapid resampling with a local order parameter presented above suggests that the early-time coarse-graining dynamics in these models \emph{effectively} results in a construction just like this, with the early-time coarse-graining length scale $\ell$ the analogue of the box size. This connection between the global information stored in an ordered state, and the local behavior of topological defects, is absent above the critical measurement rate in the gauge theory. The structured protocol shows that it can be put back in ``by hand'', so that resampling remains rapid even when an arbitrarily large fraction of spins is measured.

\noindent\textit{Discussion.---} Motivated by the challenge of efficiently extracting observables from quantum simulators,  we have developed a classical model of disruptive measurements healed by Glauber dynamics, to emulate the interplay of measurements with the local dissipative dynamics characteristic of quantum Gibbs samplers. We arrived at two striking conclusions: first, systems with local order parameters can recover or ``resample'' fast even under unstructured random protocols that measure all but a vanishing fraction of spins as $L\to \infty$. This is because topological defects are encoded in the local order parameter field, which in turn has a bias inherited from the non-measured fraction of sites. In contrast, in phases with 1-form symmetry-breaking, the relevant topological defects {\it cannot} be inferred from a local order parameter, leading to a ``resampling transition'' at sufficiently dense measurements. In this regime, the initial bias is insufficient to avoid slow resampling. However, it is possible to devise suitably structured measurement protocols that reestablish fast resampling for an arbitrarily large measured fraction. This shows that building a picture of resampling dynamics in various phases can suggest new protocols for efficiently  probing their properties.
Besides coarsening, our work also connects with the theory of error correcting codes.
In closing, we now comment on some possible extensions of the present work to more fully capture various aspects of the quantum setting. 

A quantum measurement of a local observable can instantaneously update the (conditional) state of distant degrees of freedom entangled with those being probed, while simultaneously modifying the microscopic state in the vicinity of the measurement. We have only modeled the latter; we expect that this, e.g. gives a qualitatively reasonable picture in gapped states, but not in those with critical correlations.  Finding ways to better model the entanglement-mediated modification of long-range correlations  by  measurements is an important open problem.

Going further, we might also wish to characterize rate at which one can extract {\it dynamical} correlations $\langle \mathcal{O}(\tau) \mathcal{O}(0)\rangle$ or response functions that typically require two rounds of measurements at $t=0, \tau$ with an intervening coherent evolution under $U(t) = e^{-i Ht}$. At the simplest level, one can use a classical caricature of operator spreading under unitary dynamics to generate a ``mask'' of spins corresponding to the complex, time-evolved operator $\mathcal{O}(t)$, and reset all spins within this mask; a more careful treatment could reset these spins {\it conditioned} on their joint state (rather than randomly), since this more closely resembles the effect of measuring a time-evolved local operator. Understanding resampling dynamics under such perturbations is an important step towards using quantum simulators to model real-world experiments on many-body systems.

\noindent\textit{Acknowledgments.---} We thank Akshat Pandey, John Chalker, Curt von Keyserlingk, Adam Nahum, Sarang Gopalakrishnan, and Max McGinley
for useful discussions. We  acknowledge support from UKRI Frontier Research Grant No. EP/Z002419/1 (D.T., S.A.P.). B.P. acknowledges funding through a Leverhulme-Peierls Fellowship at the University of Oxford and the Alexander von Humboldt Foundation through a Feodor-Lynen fellowship.

\bibliographystyle{apsrev4-2}
\bibliography{references}

\onecolumngrid
\vspace{2em}
\begin{center}
{\large\bfseries End Matter}
\end{center}
\twocolumngrid

\noindent\textit{Properties of Ising gauge theory.---} First, note that the IGT energy functional $\mathcal{H}_{\rm IGT}$ is invariant under an extensive number of local gauge transformations, corresponding to simultaneously flipping the spins on the six links adjacent to a single vertex. The Gibbs measure is uniform over the exponentially many states generated by this symmetry. On the dual lattice, spins sit on plaquettes, and the gauge symmetry corresponds to flipping the spins on the six faces of an elementary cube. Indeed $\mathcal{H}_{\rm IGT}$ is invariant under flipping all the spins on \emph{any} closed surface on the dual lattice. For a contractible surface, this is simply a product of gauge transformations. With periodic boundary conditions, there are additional symmetries corresponding to flipping spins on non-contractible surfaces. These are referred to as 1-form symmetries, since they act on manifolds of codimension 1. 

The model has a finite-temperature transition to a phase in which this 1-form symmetry is spontaneously broken. This type of order is referred to as topological as it is characterized only in terms of a nonlocal Wilson-loop observable, and lacks a local order parameter, i.e., all local operators exhibit exponentially decaying correlations. In 3D there are eight distinct distinct topological sectors labeled by the non-contractible $\mathbb{Z}_2$ fluxes $m_\mu\in\{0,1\}$ through the three distinct ``handles'' $\mu=x,y,z$ of the hypertorus. For example, at $T=0$ the trivial sector $(0,0,0)$ consists of all configurations obtained from the ``all up'' state by local gauge transformations, while the $(1,0,0)$ sector consists of all configurations obtained by first flipping the spins along a non-contractible dual membrane normal to the $x$-direction, and then applying local gauge transformations, and so on.

Starting from a $T=0$ state in any topological sector, flipping a spin on a single link incurs an energy cost by frustrating the four plaquettes incident on that link. These violated plaquettes form the smallest closed $\mathbb{Z}_2$ flux loop on the dual lattice. More generally, on the dual lattice, collections of flipped spins define membranes, and the flux loops are the boundaries of these membranes. Zero-temperature Glauber dynamics tries to reduce the energy by locally reducing the length of flux loops, which leads to curvature-driven dynamics. As such, in the description of coarsening, the flux loops play a role analogous to the domain walls in the Ising model, and the vortex lines in the XY model. \\

\noindent\textit{Additional information on dynamical protocol.---} The spins evolve in time under classical Glauber dynamics, which we briefly define here. At each step in the dynamics, a spin is chosen at random, and an ``update'' is proposed. The update corresponds to flipping the spin in the discrete case, and randomly re-sampling the spin from its configuration space in the continuous case. The proposed update is then accepted with probability $P_G = (1+\exp(\beta \Delta E))^{-1}$, where $\beta = 1/k_B T$ is the inverse temperature and the energy difference $\Delta E$ between the original state and the proposed new state is computed w.r.t. $\mathcal{H}$ \cite{glauber_time-dependent_1963}. A single timestep is taken to correspond to $N$ update attempts. Glauber dynamics is a detailed-balanced Markov chain, whose steady-state distribution is the Gibbs distribution of $\mathcal{H}$ at the chosen temperature $T$. In our simulations we run the Glauber dynamics at $\beta=10$: for the Ising and $\mathbb{Z}_2$ gauge cases this is essentially indistinguishable from $\beta=\infty$, whereas for the XY model this has a reduced equilibrium magnetization $m\approx 0.98$. Using a finite $\beta$ shortens the run times for the XY model, and we do not expect this to have qualitative consequences since $T\ll T_c$. 

The resampling times and logical fidelities are averaged over $10^3-10^4$ runs. In the discrete models, this requires some post-selection to account for runs which end up ``frozen'' in metastable states, rather than reaching the ground state. Such states generically appear in curvature-driven coarsening dynamics of models with discrete state spaces, and are absent in rapid-resampling regimes as $L\to \infty$ for all models we study~\cite{supmat}. Runs which end up frozen are discarded in the calculation of the average resampling time, but they are included as failed recoveries when computing the logical fidelity. \\

\noindent\textit{Rapid resampling in $O(n)$ models with extended defects.---} 
Here we extend the argument for rapid resampling in the Ising model, presented in the main text, to $O(n)$ models for $n>1$, for which the space of allowed spins is continuous rather than discrete. $O(n)$ spin models in $d$ spatial dimensions have $(d-n)$-dimensional topological defects, provided $n \leq d$. We will postpone the discussion of point defects to the Supplementary Material \cite{supmat}, and here restrict to $n < d$, so that the defects are extended. In 3D this leaves just the XY model ($n=2$), which we have shown in \autoref{fig:overview} of the main text has rapid resampling for $p<1$.

As in the Ising model, we consider first coarse-graining the initial condition (after the disruptive measurement), by averaging over blocks of spins of linear size $\ell$. For $p<1$, the magnetization of each block has a deterministic contribution from the approximately $(1-p) \ell^d$ unmeasured spins, plus a stochastic distribution from the approximately $p \ell^d$ spins which have been measured. The resulting magnetization density is
\begin{equation}
    \label{eq:vector average magnetization}
    \vec{m}(p, \ell) = (1 - p) \vec{\hat{x}} + \vec{\delta m} (p, \ell),
\end{equation}
where $\vec{\hat{x}}$ is the unit vector along which the spins were aligned before the measurement, and for $\ell$ sufficiently large $\vec{\delta m}$ can be approximated by a Gaussian random vector with length $\sim \sqrt{p / \ell^d }$. Then, provided $\ell \gg \left( \frac{p}{(1-p)^2} \right) ^{1/d}$, each block with high probability has a magnetization which is (i) of finite magnitude and (ii) aligned with that of the pre-measurement state, up to a small angular deviation $\sim \sqrt{\frac{p}{(1-p)^2 \ell^d}}$. Phrased differently, for $\ell$ sufficiently large the probability that the magnetization of a block is \emph{not} approximately aligned with the magnetization of the pre-measurement state is small. The point $p=1$ is special, since here the direction of $\vec{m}$ is everywhere uniformly random.

Our assumption is that this microscopic bias at $t=0$ leads to a bias at the level of the order parameter field at a finite time $t_\ell$, taken to be the time beyond which the description of the coarsening dynamics in terms of smooth topological defects becomes appropriate. The small probability associated with blocks of size $\ell$ that are not approximately aligned with the pre-measurement state defines a sub-critical site percolation problem at the level of blocks of spins, which bounds the size of clusters of misaligned spins: they contain $O(1)$ spins on average, while the largest such cluster contains $O(\log L)$ spins. Unlike in the Ising model, these clusters of misaligned spins cannot be thought of as minority domains, since with continuous spin variables these are not well-defined. Instead, one has to inspect the charge of the topological defects in the misaligned clusters by computing the homotopy of the direction field $\phi = \vec{m} / | \vec{m}|$ on an $(n-1)$-sphere surrounding the region \cite{mermin_topological_1979}. The bias inherited from the initial condition makes it possible to choose a sphere around each misaligned region which itself only cuts through regions which are approximately aligned with the pre-measurement magnetization, so that (i) the direction field is always well-defined since $\vec{m}$ is non-zero and (ii) $\phi$ varies only little along the sphere, guaranteeing that the homotopy is trivial. 

It follows that the net topological charge of defects contained in the clusters of misaligned spins is zero. The size of the topological defects is therefore bounded by the size of the misaligned regions, which bounds the resampling time. Just as in the Ising case, most defects decay in a finite time, while requiring that \emph{all} defects have decayed gives a resampling time $\tres \sim O(\log L)$. 

\clearpage

\begingroup
\setcounter{secnumdepth}{2}
\setcounter{figure}{0}
\setcounter{equation}{0}
\renewcommand{\thesection}{Appendix \Roman{section}}
\renewcommand{\thesubsection}{\arabic{subsection}}

\makeatletter
\@removefromreset{equation}{section}
\@removefromreset{figure}{section}
\makeatother
\renewcommand{\theequation}{S\arabic{equation}}
\renewcommand{\thefigure}{S\arabic{figure}}

\onecolumngrid

\begin{center}
	\textbf{\large Supplementary Material for ``Gibbs resampling transitions and phase-adaptive interrogation protocols''}\\[.2cm]
\end{center}


\section{\label{sec:extranumerics}Additional numerical data}

\subsection{System size dependence of resampling time}
It was argued in the main text that the dependence of resampling time on system size is either logarithmic (rapid resampling) or quadratic (slow resampling). 

\autoref{fig:resampling fast} shows the dependence of resampling time on system size in the three models considered in the main text, in their respective rapid-resampling regimes. The data is consistent with the expected logarithmic dependence on system size (though a polylogarithmic dependence cannot be ruled out from this data alone). The ``tailing off'' of the resampling time at large system size in the 3D XY model may be an artifact of the way the simulations were run (at finite temperature $\beta = 10$ and declaring resampling successful when $m > 0.98$). This is because the $\log L$ dependence comes from the requirement that \emph{all} topological defects have decayed, and requiring only the decay of an arbitrarily large fraction of the defects, which is the implication when using magnetization as a proxy, results in a finite resampling time instead.
\begin{figure}[b]
    \centering
    \includegraphics[width=\linewidth]{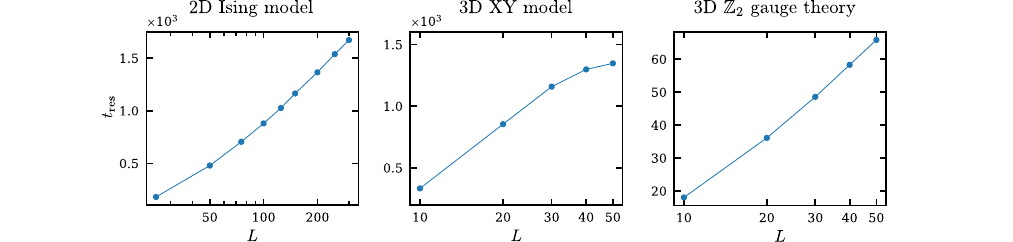}
    \caption{Resampling time as a function of system size in the 2D Ising model, the 3D XY model, and the 3D $\Z$ gauge theory, in their regimes of rapid resampling. The data was obtained at $p=0.9$ for the Ising and XY model, and at $p=0.15$ for the gauge theory.}
    \label{fig:resampling fast}
\end{figure}

\autoref{fig:resampling slow} confirms a quadratic dependence of resampling time on system size in the slow-resampling of all three models. It was obtained at measurement probability $p=1$, which is equivalent to a quench from infinite temperature into the ordered phase. For the Ising model and the XY model this is the only $p$ at which resampling is slow, and reproduces well-known results from phase-ordering kinetics \cite{bray_theory_1994}. For the 3D $\Z$ gauge theory, \autoref{fig:overview} of the main text shows that the resampling time is essentially uniform across the slow resampling phase, for $p \gtrsim 0.226$. 
\begin{figure}[t]
    \centering
    \includegraphics[width=\linewidth]{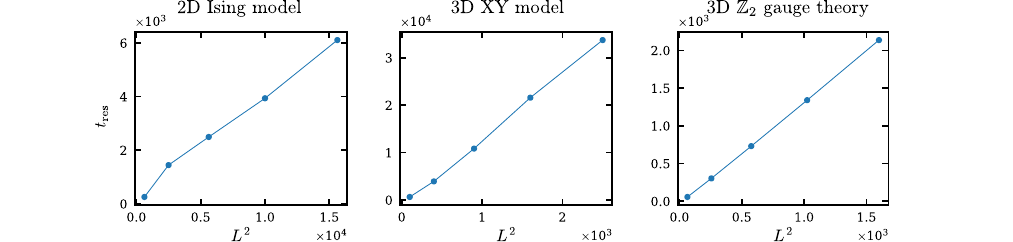}
    \caption{Resampling time as a function of quadratic system size in the 2D Ising model, the 3D XY model, and the 3D $\Z$ gauge theory, after a maximally disruptive measurement ($p=1$).}
    \label{fig:resampling slow}
\end{figure}

\subsection{Metastable states}
The state space of the Ising model and the classical $\Z$ gauge theory contains a large number of metastable states. It is possible that the Glauber dynamics, rather than reaching a ground state, end up ``frozen'' in one of the metastable states. Under zero temperature Glauber dynamics these states are absolutely stable, while at finite temperature their decay is typically very slow. Here, we will briefly discuss the structure of the metastable states in both models and show numerically that, for large enough system size, they occur only in the slow resampling regime.

In the 2D Ising model with periodic boundary conditions, the metastable states consists of parallel stripes of up and down spin domains winding one of the handles of the torus. At low but finite temperature they have a relaxation time $t \sim O(L^3 e^{4\beta})$ \cite{spirin_freezing_2001}. When the initial state is maximally random ($p=1$ in our protocol), zero temperature Glauber dynamics settles into a metastable state approximately a third of the time. (The exact probability can be obtained by computing the probabilities associated with different topologies in critical continuum percolation\cite{barros_freezing_2009}.) \autoref{fig:freezing}(a) shows the probability $M$ that a run ends up in a metastable state as a function of the measurement probability. It tends to zero at large enough system size for $p<1$, as was previously noted in Ref. \onlinecite{spirin_freezing_2001}.
\begin{figure}[b]
    \centering
    \includegraphics[width=\linewidth]{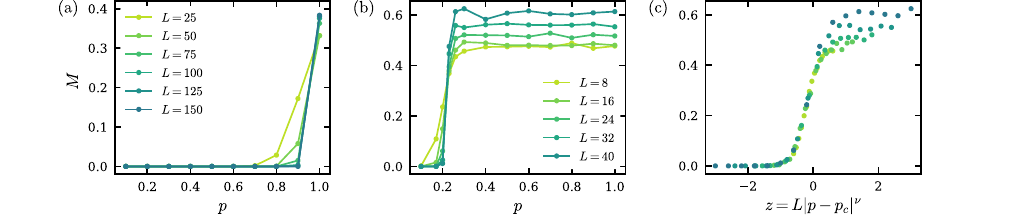}
    \caption{Rate $M$ of freezing into metastable states as a function of the measurement probability $p$, (a) in the 2D Ising model, and (b) in 3D $\Z$ gauge theory, where it undergoes a transition. (c) Scaling collapse of the freezing probability around its transition. We assume $\nu = 1$ exactly, and find $p_c \approx 0.225$, in close agreement with the threshold probability of the resampling transition.}
    \label{fig:freezing}
\end{figure}

In the 3D $\Z$ gauge theory, the metastable states consist of pairs of straight flux loops, which both non-trivially wind the same handle of the 3-torus. Such configurations are related to the ground state by flipping spins along the (dual) surface of a cylinder which is non-trivially embedded in the 3-torus. By a simple generalization of the argument given in Ref. \onlinecite{spirin_freezing_2001} from 2D to 3D, we expect the metastable states to have a relaxation time $t \sim O(L^3 \log (L)e^{4\beta})$ at low but finite temperature. \autoref{fig:freezing}(b-c) shows the probability of ``freezing'' in the 3D gauge theory as a function of $p$. The freezing probability $M$ undergoes a transition as $p$ is varied, which coincides with the resampling transition. 

In the slow resampling regime there is no bias towards any particular ground state sector, so that runs either end up frozen or, if they do not freeze, reach any of the ground states with equal probability. In the slow resampling regime, the rate of reaching metastable states therefore directly relates to the logical fidelity, as $f = (1 - M) / \rm GSD$, where GSD is the ground state degeneracy. \autoref{fig:freezing}(b) shows that $M$ increases with system size above $p_c$ in the 3D gauge theory, meaning the fidelity decreases with system size. From our results it is inconclusive whether the fidelity remains finite in the thermodynamic limit, or tends to zero as $M$ tends to 1.

\section{\label{sec:collarwidth}Collar width in the structured measurement protocol}
In all the models discussed in this work, rapid resampling is possible using a structured measurement protocol, in which the system is subdivided into ``boxes'' of linear size $R$ in which spins are measured, surrounded by a ``collar'' of unmeasured spins (\autoref{fig:collar construction}(a) of the main text). For extended defects, the required collar width $w(R)$ is set by the competition of thermal fluctuations taking defects outside of the box, and the inward pressure due to curvature. An upper bound is provided by the wandering exponent $\zeta$,  which governs fluctuations in the absence of curvature as $\Delta h \sim R^\zeta$ \cite{chaikin_principles_1995}. Here, we give a heuristic coarsening argument to suggest this bound is typically very loose.

In coarsening dynamics, defects vary on the characteristic length scale $L(t)$. As a consequence, flat defects are generically only expected on length scales \emph{smaller} than $L(t)$. For the bound $w \lesssim R^\zeta$ to be tight, a flat defect of size $\sim R$ has to form, which requires the system to evolve up to a time $t_R$ so that $L(t_R) \sim R$. During this time, however, the curvature pressure will have pushed defects away from the boundary, so that when a flat defect of size $\sim R$ is established, it will have a certain distance to the boundary of the box, constituting an ``effective collar''. Only when the $R$-dependence of this effective collar is weaker than $R^\zeta$ can we expect the wandering exponent to give a tight bound on the required collar width. 

Since $L(t)$ is the only length scale of the coarsening dynamics, the local defect velocity is $v \sim \frac{dL}{dt}$. Furthermore, as the defects are convex near the boundary of the box, this velocity is directed away from the boundary. The effective collar width at time $t_R$ is therefore proportional to 
\begin{equation}
    \int^{t_R} v \, dt \sim \int^{t_R} dL \sim R.
\end{equation}
Therefore, provided $\zeta < 1$, which is the case for all models studied in this work, for sufficiently large $R$ the fluctuations governed by the wandering exponent are smaller than the effective collar that separates flat defects from the edge of the box. Hence fluctuations \emph{outside the box} are not governed by the wandering exponent.

Instead, we expect the dominant behavior to be local thermal fluctuations, at early enough times that the effective collar is small. The rate of thermally exciting fluctuations up to a distance $w$ scales with the Boltzmann factor $\exp(-C \beta w)$, for some constant $C$. At early times, there are many defects near the boundary of the box. The relevant ``failure event'' that the collar is meant to prevent, however, is not the merging of \emph{any} defect inside the box with one from another box, but the merging of two defects from different boxes which each span their respective box, since this is the only way in which defects from more than two boxes can all merge into one, so that macroscopic defects of size $\sim L$ can form and disrupt the rapid resampling. The required collar width therefore depends on the number of places in which, at early times, a box-spanning defect is a finite distance away from the boundary. We assume that this is governed by a fractal dimension $\delta$, so that the number of boundary segments scales as $R^\delta$. Then the required width $w$ so that the probability that the spanning defect anywhere fluctuates at least a distance $w$ outside of the box is less than $\epsilon$, scales as 
\begin{equation}
    w \sim \frac{1}{\beta} \log \left( \frac{R^\delta}{\epsilon} \right).
\end{equation}
For the 2D Ising model (with $\zeta = \frac{1}{2}$), this prediction is consistent with the numerical simulation results shown in \autoref{fig:collar construction}(b) of the main text.

\section{Point defects and the 3D Heisenberg model}
Throughout this work, when presenting arguments for rapid resampling (for the uncorrelated as well as the structured protocols), we considered only models with extended defects. This excludes models like the 3D Heisenberg model, whose ``hedgehog'' defects are point-like; more generally, it excludes $O(n)$ models in $d=n$ spatial dimensions. The restriction was necessary because our arguments relied on the topological defects being confined (at zero temperature) by their geometry. By constraining the net topological charge of a finite cluster of spins to be zero, whether as a result of a local order parameter, or done explicitly as in the structured measurement protocol, extended defects are guaranteed to form closed loops (or spheres, etc.) which are entirely contained within the cluster. In the dynamics, this guarantees an \emph{inward} curvature pressure which shrinks the defect. Point defects are not geometrically confined, but instead confined only by their interaction: pairs with equal charge repel, and pairs with opposite charge attract. This is not strictly sufficient to guarantee that defects remain inside the region they were created in, even at zero temperature: if a defect in one cluster is closer to a defect of opposite charge in a neighboring cluster than it is to its ``partner'' in its own cluster, then the defect will be pulled \emph{outward} rather than in.

The interaction between defects in $O(n)$ models is very strong ($F(r) \sim r^{d-3}$ \cite{bray_theory_1994}), and furthermore the local order parameter (or the structured measurement protocol) guarantees charge neutrality on a finite length scale. We expect that these two things are often sufficient to guarantee fast phase ordering. Indeed, for the Heisenberg model, we find numerically that resampling is rapid for $p<1$ (\autoref{fig:Heisenberg}). However, we are not aware of a general argument, and therefore leave the case of point defects as a conjecture.
\begin{figure}
    \centering
    \includegraphics[width=\linewidth]{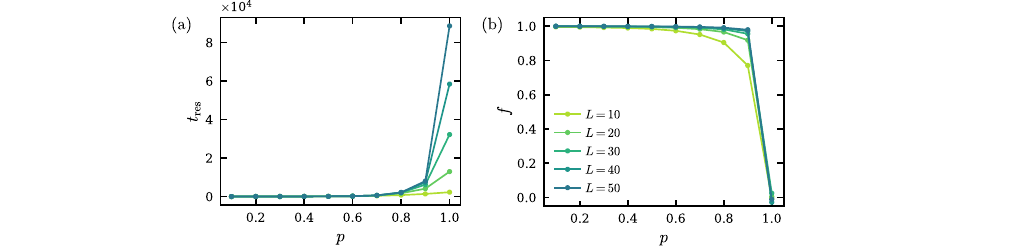}
    \caption{(a) Resampling time and (b) logical fidelity as a function of measurement probability $p$ in the 3D Heisenberg model. The fidelity is defined using the root-mean-squared angular deviation of the final magnetization vector from the pre-measurement magnetization axis, as $f = 1 - \frac{2 \theta_{\rm{rms}}}{\pi^2 - 4}$ (normalized to the uniform distribution to give zero at $p=1$).}
    \label{fig:Heisenberg}
\end{figure}

\endgroup

\end{document}